\RequirePackage{fix-cm}
\documentclass[twocolumn,epjc3]{svjour3}
\smartqed  
\RequirePackage{graphicx}
\usepackage{amssymb}
\journalname{Eur. Phys. J. C}
\begin{document}

\title{Comparing accretion disk profiles of Bogush-Galt'sov naked singularity and Kerr black hole}


\author{R.Kh. Karimov\thanksref{e1,addr1}
        \and
        R.N. Izmailov\thanksref{e2,addr1}
        \and
        A.A. Potapov\thanksref{e3,addr2}
        \and
        K.K. Nandi\thanksref{e4,addr1}
}

\thankstext{e1}{e-mail: karimov\_ramis\_92@mail.ru}
\thankstext{e2}{e-mail: izmailov.ramil@gmail.com}
\thankstext{e3}{e-mail: a.a.potapov@strbsu.ru}
\thankstext{e4}{e-mail: kamalnandi1952@rediffmail.com}


\institute{Zel'dovich International Center for Astrophysics, M. Akmullah Bashkir State Pedagogical University, 3A,
           October Revolution Street, Ufa 450008, RB, Russia \label{addr1}
           \and
           Department of Physics \& Astronomy, Bashkir State University, 47A, Lenin Street, Sterlitamak 453103, RB, Russia \label{addr2}
}

\date{Received: date / Accepted: date}

\maketitle

\begin{abstract}
It is well known that the Einstein-scalar system of general relativity can in principle yield non-unique exact spinning naked singularities, which lead to unique Kerr black hole when the scalar field is switched off. It is a challenging task to observationally distinguish these two types of objects. Since accretion process could be a viable diagnostic for this distinction, the purpose of the present work is to explore whether there could be features in the accretion profiles distinguishing the singularity from a Kerr black hole. Here we study the Novikov-Thorne thin accretion to a \textit{new} spinning naked singularity with a scalar charge $\sigma$ recently reported by Bogush and Gal'tsov (BG). Our study reveals that: (1) The conversion efficiency $\epsilon$ of the BG naked singularity is \textit{independent} of $\sigma$ and (2) The maxima of emissivity profiles for the BG singularity tend to shift towards the inner disk ISCO boundary $r=r_{\scriptsize{\textmd{ms}}}$ and peak at a value significantly larger than those of a Kerr black hole with the increase of $a$, $\sigma$ and relative shrinking of $\sqrt{-g}$. All these effects are \textit{quantitatively} tabulated, which reveal, for instance, that the flux from the naked singularity could be as high as $10^{5}$ times larger than that of a Kerr black hole. Since these distinguishing features are known to be shared also by other models of naked singularity, it is tempting to speculate that such behavior could be hallmark of naked singularities.
\end{abstract}

\section{Introduction}
\label{intro}
After the advent of Brans-Dicke (BD) theory, originally formulated in the Jordan Frame (JF) as a Machian competitor to Einstein's general relativity, scalar fields have become a significant component of research in gravitational physics today. Solar system experiments suggest that the dimensionless BD coupling constant $\omega_{\scriptsize{\textmd{BD}}}$ should be very large, $\left\vert\omega_{\scriptsize{\textmd{BD}}}\right\vert\geq 50,000$, and in this limit the two theories become practically identical (however, see \cite{Bhadra:2001a,Faraoni:1999}). In the conformally rescaled Einstein Frame (EF), the original JF formulation becomes Einstein's general relativity (GR), which has found many applications including the interpretation of galactic halo (see, e.g., \cite{Matos:2000,Nandi:2009}). Applying the generation technique devised in \cite{Tiwari:1976,Singh:1979} together with a radial redefinition proposed in \cite{Misra:1972}, Kim \cite{Kim:1999} generated a BD-Kerr-Newman type naked singularity solution. Its chargeless corollary is the BD-Kerr type naked singularity for which the steady state Novikov-Thorne \cite{Novikov:1973,Shakura:1973,Page:1974,Thorne:1974} accretion disk emissivity properties have been recently studied by Shahidi, Harko and Kov\'{a}cs (SHK) \cite{Shahidi:2020}. On the other hand, a new naked singularity solution in the Einstein-scalar theory has been obtained recently by Bogush and Galt'sov (BG) \cite{Bogush:2020} that yield Kerr black hole in the zero scalar field limit. Since the signatures of accretion to naked singularity could be an important diagnostic distinguishing it from black hole, our motivation here is to precisely explore those distinguishing features in the comparative accretion profiles of the singularity and Kerr black hole. With this in mind, we shall study the accretion properties of the BG solution \cite{Bogush:2020}.

In general, due to the existence of non-trivial scalar field (often called hair) either in the JF or in the EF, the solutions might show singularity at a finite radius, where curvature invariants diverge. It is called a naked singularity since nothing in spacetime prevents light to reach external asymptotic observers thereby making it visible. Penrose's cosmic censorship conjecture \cite{Penrose:1969} forbids occurrence of naked singularity (strong version) or if it occurs, it must be clothed by an event horizon (weak version) preventing light to reach asymptotic observers. There is also the Ruffini-Wheeler conjecture \cite{Ruffini:1971} that states that "black holes can have no hair". Since the two conjectures are not yet rigorously proven, the final fate of gravitational collapse does not rule out naked singularities, hairy or otherwise \cite{Joshi:2007}. A notable realistic scenario involves perfect fluid collapse which, depending on the initial conditions, can lead to a naked singularity known as the JMN solution \cite{Joshi:2014}. Early static singular solution with scalar hair in the EF is well known \cite{Janis:1968}. (See also, \cite{Bhadra:2001b}). There is a large literature on scalar field solutions (not necessarily singular ones) both in JF and EF including works on BD wormholes threaded by exotic scalar field \cite{Agnese:1995,Nandi:1997,Anchordoqui:1997,Nandi:1998,Nandi:2004,Bhattacharya:2009,Cardoso:2016}. Some works relate to the possible detectability of wormholes and naked singularities by their gravitational lensing signatures \cite{Nandi:2006,Nandi:2017,Lukmanova:2018,Virbhadra:2002,Izmailov:2019,Tsukamoto:2017,Shaikh:2017,Nakajima:2012,Shaikh:2019,Shaikh:2018} and by their accretion disk characteristics \cite{Longair:1994,Torres:2002,Bambi:2013,Karimov:2019,Harko:2009a,Harko:2009b,Karimov:2020,Yusupova:2021,Heydari-Fard:2021}. There are important works related to material collapse to black holes - an early one being the famous hoop conjecture due to Thorne \cite{Thorne:1972} and its recent inverse due to Hod \cite{Hod:2018,Hod:2020a,Hod:2020b,Nandi:2020,Nandi:2021}, stability issues in generic scalar tensor theory (see, e.g., \cite{Bronnikov:2011}) - the list is by no means exhaustive. There is however a useful introductory review on this topic in SHK \cite{Shahidi:2020} and comprehensive references are available therein.

If naked singularities truly exist in nature, it should have some distinct observable signatures. Two important observational diagnostics come to mind. One is to measure their accretion emissivity profiles to see how they differ from those of Kerr black hole and the other is to measure their shadows projected on the background of the corresponding accretion flows. We shall focus on the former diagnostic in this paper. SHK \cite{Shahidi:2020} studied this diagnostic for BD-Kerr type naked singularity from which, depending on the ranges of the solution parameter $\gamma$ (related to $\omega_{\scriptsize{\textmd{BD}}}$), BD-Kerr type hairy black hole also followed. The occurrence of scalar hairy black holes can have implications for Penrose censorship conjecture as discussed in \cite{Shahidi:2020}. In contrast, no hairy black holes follow from the BG naked singularity of GR \cite{Bogush:2020} reflecting the difference between the BD and GR black holes.

In this paper, we study thin accretion disk radiation profiles of the BG solution using the celebrated Novikov-Thorne model, ingredients of which we briefly summarize for easy reading in Sec.2. Next, in Sec.3, assuming a toy model of stellar sized BG spinning naked singularity, we compute the emissivity profiles (Sec.4), and compare them with those of Kerr black hole. Sec.5 summarizes our findings. We shall take units such that $8\pi G=c=1$.

\section{Thin accretion disk: Novikov-Thorne model in brief}
\label{sec:2}
For our purposes, we start with the general form of the spinning metric
\begin{equation}
d\tau^{2} = g_{tt}dt^{2} + 2g_{t\varphi}dtd\varphi + g_{rr}dr^{2} + g_{\theta\theta}d\theta^{2}+ g_{\varphi\varphi}d\varphi^{2},
\label{Eq1}
\end{equation}%
and from it find two sets of formulas related to the accretion disk. These are:

\subsection{Kinematic formulas}

Since the BG metric components are independent of $t$ and $\varphi$, there exist conserved quantities, which are the specific energy $\widetilde{E}$, the specific angular momentum $\widetilde{L}$ of test particles moving in stable orbits. For particles moving on the equatorial plane ($\theta = \pi/2$), they are
\begin{eqnarray}
g_{tt} \frac{dt}{d\tau} + g_{t\varphi} \frac{d\varphi}{d\tau} &=& -\widetilde{E}, \\
g_{t\varphi} \frac{dt}{d\tau} + g_{\varphi\varphi} \frac{d\varphi}{d\tau} &=& \widetilde{L}.
\end{eqnarray}%
Solving, one obtains
\begin{eqnarray}
\frac{dt}{d\tau} &=& \frac{\widetilde{E} g_{\varphi\varphi} + \widetilde{L}g_{t\varphi }}{g_{t\varphi}^{2} - g_{tt}g_{\varphi\varphi}}, \\
\frac{d\varphi}{d\tau} &=& -\frac{\widetilde{E} g_{t\varphi} + \widetilde{L}g_{tt}}{g_{t\varphi}^{2} - g_{tt}g_{\varphi\varphi}}.
\end{eqnarray}%
From the metric, then it follows for massive test particles that \cite{Harko:2009a}
\begin{equation}
g_{rr}\left(\frac{dr}{d\tau}\right)^{2} = V_{\scriptsize{\textmd{eff}}}(r; \widetilde{E}, \widetilde{L}).
\end{equation}%
The last equation thus provides an effective potential term
\begin{equation}
V_{\scriptsize{\textmd{eff}}}(r;\widetilde{E},\widetilde{L}) = -1 + \frac{\widetilde{E}^{2} g_{\varphi\varphi} + 2\widetilde{E}\widetilde{L} g_{t\varphi} + \widetilde{L}^{2} g_{tt}}{g_{t\varphi}^{2} - g_{tt}g_{\varphi\varphi}}.
\end{equation}%
Circular orbit radii in the equatorial plane follow from $V_{\scriptsize{\textmd{eff}}}\left(r\right) = 0$ and $V_{\scriptsize{\textmd{eff}}, r}(r) = 0$, where the comma in the subscript denotes a derivative with respect to the radial coordinate $r$. These conditions allow us to write the kinematic quantities as
\begin{eqnarray}
\widetilde{E} &=& -\frac{g_{tt} + g_{t\varphi}\Omega}{\sqrt{-g_{tt} - 2g_{t\varphi}\Omega - g_{\varphi\varphi}\Omega^{2}}}, \\
\widetilde{L} &=& \frac{g_{t\varphi} +g_{\varphi\varphi}\Omega}{\sqrt{-g_{tt} -2g_{t\varphi}\Omega -g_{\varphi\varphi}\Omega^{2}}}, \\
\Omega &=& \frac{d\varphi}{dt} = -\frac{-g_{t\varphi, r}\pm \sqrt{\left(g_{t\varphi, r}\right)^{2} -g_{tt, r}g_{\varphi\varphi, r}}}{g_{\varphi\varphi, r}}.
\end{eqnarray}

Following Novikov-Thorne model \cite{Novikov:1973}, an important characteristic of the accretion disk is the \textit{efficiency} $\epsilon$, which quantifies the ability by which the accreting body converts particle mass into radiation. It is measured at infinity and is defined as the ratio between the rate of energy of the photons emitted from the disk surface and the rate with which mass-energy is transported to the central accreting body. If all photons reach asymptotic infinity, the efficiency is given by the specific energy $\widetilde{E}$ of accreting particles measured at $r=r_{\scriptsize{\textmd{ms}}}$ such that \cite{Page:1974,Thorne:1974}
\begin{equation}
\epsilon = 1 - \widetilde{E}_{\scriptsize{\textmd{ms}}}.
\end{equation}
The behavior of the potential $V$ in Eq.(7) gives \textit{marginally stable} circular radius $r=r_{\scriptsize{\textmd{ms}}}$ defining an inner edge or innermost stable circular orbit (ISCO) radius as a the solution of
\begin{equation}
\left. d^{2}V_{\scriptsize{\textmd{eff}}}/dr^{2}\right\vert_{r=r_{\scriptsize{\textmd{ms}}}} = 0,
\end{equation}%
while the orbits at higher radii are Keplerian. We emphasize that the above kinematic formulas $\widetilde{E}$, $\widetilde{L}$, $\Omega$, $\epsilon$ and $V_{\scriptsize{\textmd{eff}}}$ depend only on $g_{tt}$, $g_{\varphi\varphi}$ and $g_{t\varphi}$, which are the same as the Kerr metric and therefore the $r_{\scriptsize{\textmd{ms}}}$ exactly coincide with those for the Kerr geometry obtained long ago by Bardeen, Press and Teukolsky \cite{Bardeen:1972}. So we omit displaying the well known kinematic plots for Kerr black hole but the formulas will anyway be needed for emissivity profile computations in the sequel.

\subsection{Emissivity formulas }

We assume the accretion disk to be thin such that $\left\vert\theta - \pi/2 \right\vert \ll 1$ and the height $H$ much smaller than the characteristic radius $R$ of the disk, $H\ll R$. The disk is assumed to be in hydrodynamical equilibrium stabilizing its vertical size, with the pressure and vertical entropy gradient being negligible in the disk. The efficient cooling via the radiation over the disk surface is assumed to be preventing the disk from collecting the heat generated by stresses and dynamical friction. In steady-state accretion disk models, the mass accretion rate $\dot{M}_{0}$ is assumed to be constant and the physical quantities describing the orbiting matter are averaged over a characteristic time scale, e.g., the total period of the orbits over the azimuthal angle $\Delta\varphi = 2\pi$ , and over the height $H$.

In the above steady-state thin disk model, the orbiting particles have $\Omega$ , $\widetilde{E}$ and $\widetilde{L}$ that depend only on the radii of the orbits. Accreting particles orbiting with the four-velocity $u^{\mu}$ form a disk of an averaged surface density $\rho$. Novikov and Thorne \cite{Novikov:1973}, using the rest mass conservation law, showed that the time averaged rate of rest mass accretion $dM_{0}/dt$ is independent of the radius, i.e., $\dot{M_{0}} \equiv dM_{0}/dt = -2\pi ru^{r}\rho =$ \textmd{const}. (Here $u^{r}$ is the radial component of the four-velocity). We omit other known technical details of the model but quote only the relevant formulas below.

These consist of the flux $F(r)$ , temperature $T(r)$ and the lumininosity of the radiant energy $L\left( \nu \right)$ over the disk that can be expressed in terms of $\Omega$, $\widetilde{E}$ and $\widetilde{L}$ of the compact sphere as
\begin{equation}
F\left( r \right) = - \frac{\dot{M}_{0}}{4\pi\sqrt{-g}} \frac{\Omega_{, r}}{\left(\widetilde{E} - \Omega\widetilde{L}\right)^{2}} \int_{r_{\scriptsize{\textmd{ms}}}}^{r} \left(\widetilde{E} - \Omega\widetilde{L}\right) \widetilde{L}_{, r}dr.
\end{equation}%
Because of the assumed thermodynamical equilibrium, the radiation flux emitted by the disk surface will follow Stefan-Boltzmann law:%
\begin{equation}
F\left( r\right) = \sigma_{0}T^{4}\left( r \right) ,
\end{equation}%
where $\sigma_{0}$ is the Stefan-Boltzmann constant. The observed luminosity $L\left( \nu \right)$ has a redshifted black body spectrum
\begin{equation}
L\left( \nu \right) = 4\pi\textmd{d}^{2}I(\nu) = \frac{8\pi h\cos{j}}{c^{2}}%
\int_{r_{\scriptsize{\textmd{ms}}}}^{r_{\scriptsize{\textmd{f}}}}\int_{0}^{2\pi }\frac{\nu_{e}^{3}rdrd%
\varphi }{\exp \left[ \frac{h\nu _{e}}{k_{B}T}\right] -1}.
\end{equation}%
Here $h$ is Planck's constant, $I\left(\nu\right)$ is the Planck distribution function, $k_{B}$ is the Boltzmann constant, $\nu_{e}$ is the emission frequency, d is the distance to the source, $j$ is the disk inclination angle perpendicular to the line of sight, and $r_{\scriptsize{\textmd{ms}}}$ and $r_{\scriptsize{\textmd{f}}}$ indicate the radii of the innermost and outermost edge of the disk, respectively. We take $r_{\scriptsize{\textmd{f}}} \rightarrow \infty$, since we expect that the flux over the disk surface should vanish at $r \rightarrow\infty$ due to the assumed asymptotic flatness of the geometry. We assume $j = 0^{\circ}$ so that the disk is face-on. The observed photons are redshifted to the reception frequency $\nu_{r}$ related to the emission frequency $\nu_{e}$ by
\begin{equation}
\nu_{e} = (1+z) \nu_{r},
\end{equation}%
where the red-shift factor is given by \cite{Harko:2009a}
\begin{equation}
1+z = \frac{1+\Omega r \sin{\varphi} \sin{j}}{\sqrt{-g_{tt} - 2g_{t\varphi}\Omega - g_{\varphi\varphi}\Omega^{2}}}.
\end{equation}%
We shall numerically integrate out Eq.(15) over $r$ and $\varphi$, taking into account the expressions of redshift factor from Eqs. (16) and (17), to compute the luminosity $L\left( \nu \right) $ as a function of frequency $\nu$. It is important to note that the denominator in Eq.(17) becomes smaller as $r \rightarrow r_{\scriptsize{\textmd{ms}}}$, hence the redshift factor ($1+z$) becomes large causing a large luminosity $\nu L\left( \nu \right)$ vs $\nu$ [Hz] plotted in Figs.3, whereas the light bending effect becomes negligible compared to it (see, e.g., \cite{Shahidi:2020}) and hence ignored in the present work.

\section{Comparison of emissivity profiles of BG spinning naked singularity and Kerr black hole}
\label{sec:3}

\subsection{Bogush-Galt'sov (BG) spinning naked singularity}

Einstein field equations coupled minimally to a scalar field $\phi$ considered by BG are \cite{Bogush:2020}:
\begin{eqnarray}
R_{\mu\nu} &=& 2\partial_{\mu}\phi \partial_{\nu}\phi, \\
\mathit{\nabla}_{\mu}\mathit{\nabla}^{\mu}\phi &=& 0.
\end{eqnarray}%
The complete solution is given in the form
\begin{eqnarray}
d\tau_{\scriptsize{\textmd{BG}}}^{2} &=& -f\left(dt-\omega d\varphi\right)^{2} + f^{-1}h_{ij}dx^{i}dx^{j}, \\
f(r, \theta) &=& \frac{\Delta - a^{2}\sin^{2}{\theta}}{r^{2} + a^{2}\cos^{2}{\theta}}, \\
\omega (r, \theta) &=& -\frac{2aMr\sin^{2}\theta}{\Delta - a^{2}\sin^{2}{\theta}}, \\
h_{ij}dx^{i}dx^{j} &=& H(r, \theta) \left(dr^{2} + \Delta d\theta^{2}\right) + \Delta \sin^{2}{\theta} d\varphi^{2}, \\
\Delta &=& (r-M)^{2}-b^{2}, \\
b &=& \sqrt{M^{2}-a^{2}},
\end{eqnarray}%
with the scalar field $\phi$ and the function $H$ given by
\begin{eqnarray}
\phi &=& \phi_{\infty} + \frac{\sigma}{2b} \ln \left(\frac{r-M+b}{r-M-b}\right) , \\
H(r,\theta ) &=&\left( \frac{\Delta -a^{2}\sin ^{2}\theta }{\Delta }\right) \left[ 1+\frac{b^{2}}{\Delta }\sin ^{2}\theta \right] ^{-\sigma ^{2}/b^{2}}.
\end{eqnarray}%
where $M$ is the mass, $a$ is the spin and $\sigma $ is called the scalar charge\footnote{%
BG \cite{Bogush:2020} denote the scalar charge by an unusual notation $\Sigma$. This notation is customarily used in the literature to denote the expression $\Sigma = r^{2} + a^{2}\cos^{2}{\theta}$. To avoid any confusion, we have in this paper replaced the BG scalar charge $\Sigma$ by $\sigma$. Also, $\omega (r, \theta)$ of the BG solution should not be confused with the BD coupling constant $\omega_{\scriptsize{\textmd{BD}}}$.}.

It has also received independent support from the literature. For instance, Chauvineau \cite{Chavineau:2019} obtained a spinning solution, which coincides with the BG solution in the same extreme limit $b^{2} = M^{2}-a^{2}\rightarrow 0$. The curvature scalar $R$ for the BG solution diverges at $\Delta = 0$, $b^{2}\geq 0$ yielding a singular radius
\begin{equation}
r_{\scriptsize{\textmd{S}}} = M + \sqrt{M^{2} - a^{2}} = M\left(1+\sqrt{1-a_{\ast}^{2}}%
\right),
\end{equation}%
where $a_{\ast} = a/M$. That's why we call the BG solution a spinning naked singularity, not covered by event horizon. The solution has ring singularities in the equatorial plane ($\theta =\pi /2$) at $r=0$ for any $b$ and at $r=M$ for $b^{2}<0$. In the Kerr case ($\sigma =0$), the condition $\Delta = 0$ yields event horizon at the same radius, $r_{\scriptsize{\textmd{H}}} = M+\sqrt{M^{2}-a^{2}}$. We emphasize that BG spinning hairy ($\sigma \neq 0$) naked singularity for which $a_{\ast}<1$ is \textit{not} the hairless usual Kerr naked singularity for which $a_{\ast}>1$. If the condition $a_{\ast}>1$ is imposed on the BG solution, then $r_{\scriptsize{\textmd{S}}}$ becomes imaginary and the solution would \textit{look like} a Kerr naked singularity but with hair $\sigma$. It would be certainly very interesting to study the spacetime structures of hairy Kerr-like ($\sigma \neq 0$) and hairless Kerr ($\sigma = 0$) naked singularities both now defined by $a_{\ast }>1$. While work on this topic is underway, in this paper we focus only on the distinctive accretion properties between BG naked singularity and Kerr black hole.

\subsection{Emissivity maxima}

Since kinematic properties of the BG accretion disk are exactly the same as those of Kerr, we shall consider only the emissivity properties of both objects. For this, we shall assume a toy model of a stellar sized accreting object to be $M=15M_{\odot}$ with an accretion rate $\dot{M}_{0} = 10^{18}$ gm.sec$^{-1}$. The inner boundary of accretion is determined by the ISCO radius $r_{\scriptsize{\textmd{ms}}} > r_{\scriptsize{\textmd{S}}}$, which, for the BG solution, is \textit{independent} of the scalar charge $\sigma $ but depends on the spin $a$, as accurately computed in Table 1. We adopt the Thorne black hole spin limit $a/M=0.998$ for BG naked singularity as well to make the comparison with Kerr black hole meaningful. Moreover, Eq.(28) suggests a \textit{real valued} singular radius $r_{\scriptsize{\textmd{S}}}$ \ only for $a/M\leq 1$, which for the Kerr black hole exactly coincides with the horizon radius. The BG naked singularity is very much unlike the Kerr naked singularity that occurs at $a/M>1$ rendering the horizon radius $r_{\scriptsize{\textmd{H}}}$ \textit{imaginary} or non-existent. There is no limit on the scalar charge $\sigma$, but for definiteness of calculations, we shall always restrict it as $\sigma /M\leq 1 $. The Kerr kinematic values displayed in the table will be needed for numerical computation of the emissivity properties of the BG naked singularity.

\begin{table}[!ht]
\caption{The $r_{\textmd{\scriptsize{ms}}}$, $r_{\scriptsize{\textmd{S}}}$, $\tilde{E}_{\scriptsize{\textmd{ms}}}$ and the efficiency $\epsilon$ for BG NS. They are the same as those of Kerr BH. The general relativistic Schwarzschild black hole corresponds to $a=0$.}
\centering
\begin{tabular}{|c|c|c|c|c|}
\hline
$a/M$ & $r_{\scriptsize{\textmd{ms}}}$/$M$ & $r_{\scriptsize{\textmd{S}}}$/$M$ & $\tilde{E}_{\scriptsize{\textmd{ms}}}$ &
$\epsilon [\%]$ \\ \hline
$0$ & $6$ & $2$ & $0.943$ & $5.7$ \\
$0.1$ & $5.6693$ & $1.9949$ & $0.939$ & $6.1$ \\
$0.3$ & $4.9786$ & $1.9539$ & $0.931$ & $6.9$ \\
$0.5$ & $4.2330$ & $1.8660$ & $0.918$ & $8.2$ \\
$0.7$ & $3.3931$ & $1.7141$ & $0.896$ & $10.4$ \\
$0.9$ & $2.3209$ & $1.4358$ & $0.844$ & $15.6$ \\
$0.95$ & $1.9372$ & $1.3122$ & $0.810$ & $19.0$ \\
$0.998$ & $1.2370$ & $1.0632$ & $0.679$ & $32.1$ \\ \hline
\end{tabular}
\end{table}

\begin{table*}[!ht]
\caption{The exact maximum values of the time averaged radiation flux $F_{\scriptsize{\textmd{max}}}(r)$, temperature $T_{\scriptsize{\textmd{max}}}(r)$ and the emission spectra $\nu L(\nu)_{\scriptsize{\textmd{max}}}$ for accretion disk around BG naked singularity. The table shows the Kerr maximum values ($\sigma =0$) in the first column for comparison.}
\centering
\begin{tabular}{|c|c|c|c|c|}
\hline
\multicolumn{5}{|c|}{$a=0$} \\ \hline
$\sigma /M$ & $0$ & $0.3$ & $0.7$ & $1$ \\ \hline
$F_{\scriptsize{\textmd{max}}}(r)[\textmd{erg}\cdot \textmd{s}^{-1}\cdot \textmd{cm}^{-2}]$ &
$5.739\times 10^{14}$ & $5.748\times 10^{14}$ & $5.787\times 10^{14}$ & $5.838\times 10^{14}$ \\ \hline
$T_{\scriptsize{\textmd{max}}}(r)[K]$ & $5.640\times 10^{4}$ & $5.642\times 10^{4}$ & $5.652\times 10^{4}$ & $5.664\times 10^{4}$ \\ \hline
$\nu L(\nu)_{\scriptsize{\textmd{max}}}[\textmd{erg}\cdot \textmd{s}^{-1}]$ &
$4.771\times 10^{30}$ & $4.774\times 10^{30}$ & $4.787\times 10^{30}$ & $4.804\times 10^{30}$ \\
\hline\hline
\multicolumn{5}{|c|}{$a=0.3M$} \\ \hline
$\sigma /M$ & $0$ & $0.3$ & $0.7$ & $1$ \\ \hline
$F_{\scriptsize{\textmd{max}}}(r)[\textmd{erg}\cdot \textmd{s}^{-1}\cdot \textmd{cm}^{-2}]$ &
$1.275\times 10^{15}$ & $1.278\times 10^{15}$ & $1.292\times 10^{15}$ & $1.310\times 10^{15}$ \\ \hline
$T_{\scriptsize{\textmd{max}}}(r)[K]$ & $6.887\times 10^{4}$ & $6.891\times 10^{4}$ & $6.909\times 10^{4}$ & $6.933\times 10^{4}$ \\ \hline
$\nu L(\nu)_{\scriptsize{\textmd{max}}}[\textmd{erg}\cdot \textmd{s}^{-1}]$ &
$6.961\times 10^{30}$ & $6.967\times 10^{30}$ & $6.996\times 10^{30}$ & $7.034\times 10^{30}$ \\
\hline\hline
\multicolumn{5}{|c|}{$a=0.7M$} \\ \hline
$\sigma /M$ & $0$ & $0.3$ & $0.7$ & $1$ \\ \hline
$F_{\scriptsize{\textmd{max}}}(r)[\textmd{erg}\cdot \textmd{s}^{-1}\cdot \textmd{cm}^{-2}]$ &
$6.833\times 10^{15}$ & $6.876\times 10^{15}$ & $7.076\times 10^{15}$ & $7.343\times 10^{15}$ \\ \hline
$T_{\scriptsize{\textmd{max}}}(r)[K]$ & $1.048\times 10^{5}$ & $1.049\times 10^{5}$ & $1.057\times 10^{5}$ & $1.067\times 10^{5}$ \\ \hline
$\nu L(\nu)_{\scriptsize{\textmd{max}}}[\textmd{erg}\cdot \textmd{s}^{-1}]$ &
$1.512\times 10^{31}$ & $1.516\times 10^{31}$ & $1.532\times 10^{31}$ & $1.553\times 10^{31}$ \\
\hline\hline
\multicolumn{5}{|c|}{$a=0.9M$} \\ \hline
$\sigma /M$ & $0$ & $0.3$ & $0.7$ & $1$ \\ \hline
$F_{\scriptsize{\textmd{max}}}(r)[\textmd{erg}\cdot \textmd{s}^{-1}\cdot \textmd{cm}^{-2}]$ &
$3.908\times 10^{16}$ & $3.985\times 10^{16}$ & $4.352\times 10^{16}$ & $4.899\times 10^{16}$ \\ \hline
$T_{\scriptsize{\textmd{max}}}(r)[K]$ & $1.620\times 10^{5}$ & $1.628\times 10^{5}$ & $1.664\times 10^{5}$ & $1.714\times 10^{5}$ \\ \hline
$\nu L(\nu)_{\scriptsize{\textmd{max}}}[\textmd{erg}\cdot \textmd{s}^{-1}]$ &
$3.253\times 10^{31}$ & $3.274\times 10^{31}$ & $3.374\times 10^{31}$ & $3.513\times 10^{31}$ \\
\hline\hline
\multicolumn{5}{|c|}{$a=0.95M$} \\ \hline
$\sigma /M$ & $0$ & $0.3$ & $0.7$ & $1$ \\ \hline
$F_{\scriptsize{\textmd{max}}}(r)[\textmd{erg}\cdot \textmd{s}^{-1}\cdot \textmd{cm}^{-2}]$ &
$9.417\times 10^{16}$ & $9.760\times 10^{16}$ & $1.155\times 10^{17}$ & $1.468\times 10^{17}$ \\ \hline
$T_{\scriptsize{\textmd{max}}}(r)[K]$ & $2.019\times 10^{5}$ & $2.037\times 10^{5}$ & $2.124\times 10^{5}$ & $2.256\times 10^{5}$ \\ \hline
$\nu L(\nu)_{\scriptsize{\textmd{max}}}[\textmd{erg}\cdot \textmd{s}^{-1}]$ &
$4.665\times 10^{31}$ & $4.719\times 10^{31}$ & $4.978\times 10^{31}$ & $5.364\times 10^{31}$ \\
\hline\hline
\multicolumn{5}{|c|}{$a=0.998M$} \\ \hline
$\sigma /M$ & $0$ & $0.3$ & $0.7$ & $1$ \\ \hline
$F_{\scriptsize{\textmd{max}}}(r)[\textmd{erg}\cdot \textmd{s}^{-1}\cdot \textmd{cm}^{-2}]$ &
$1.176\times 10^{18}$ & $1.798\times 10^{18}$ & $1.721\times 10^{20}$ & $5.267\times 10^{23}$ \\ \hline
$T_{\scriptsize{\textmd{max}}}(r)[K]$ & $3.795\times 10^{5}$ & $4.219\times 10^{5}$ & $1.319\times 10^{6}$ & $9.817\times 10^{6}$ \\ \hline
$\nu L(\nu)_{\scriptsize{\textmd{max}}}[\textmd{erg}\cdot \textmd{s}^{-1}]$ &
$1.082\times 10^{32}$ & $1.166\times 10^{32}$ & $3.685\times 10^{32}$ & $1.789\times 10^{35}$ \\
\hline
\end{tabular}
\end{table*}

\begin{figure*}[!ht]
  \centerline{\includegraphics[scale=0.6]{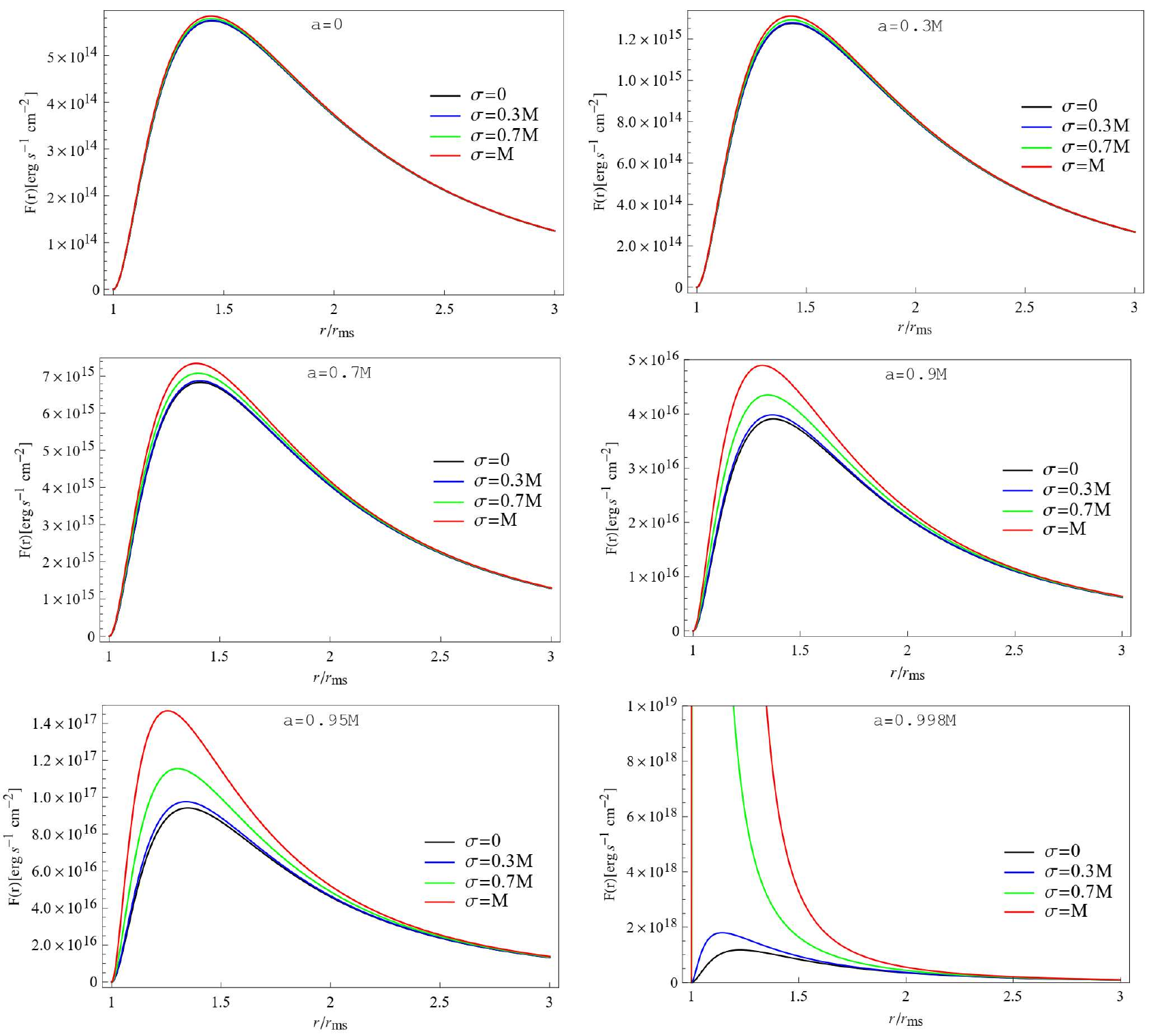}}
  \caption{Variation of radiative flux with distances away from $r_{\scriptsize{\textmd{ms}}}$.}
  \label{Flux}
\end{figure*}

\begin{figure*}[!ht]
  \centerline{\includegraphics[scale=0.6]{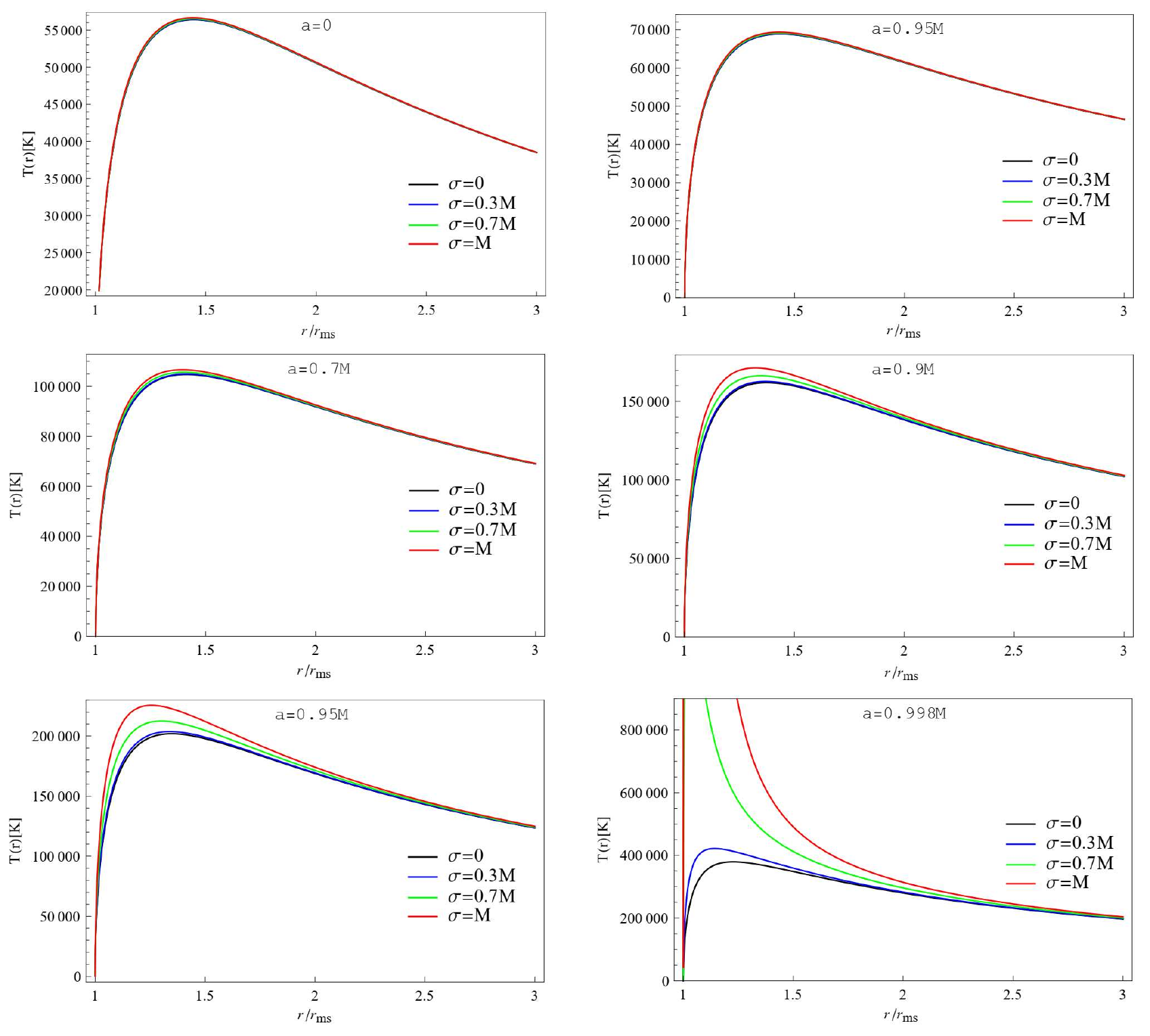}}
  \caption{Variation of temperature with distances away from $r_{\scriptsize{\textmd{ms}}}$.}
  \label{Temp}
\end{figure*}

\begin{figure*}[!ht]
  \centerline{\includegraphics[scale=0.6]{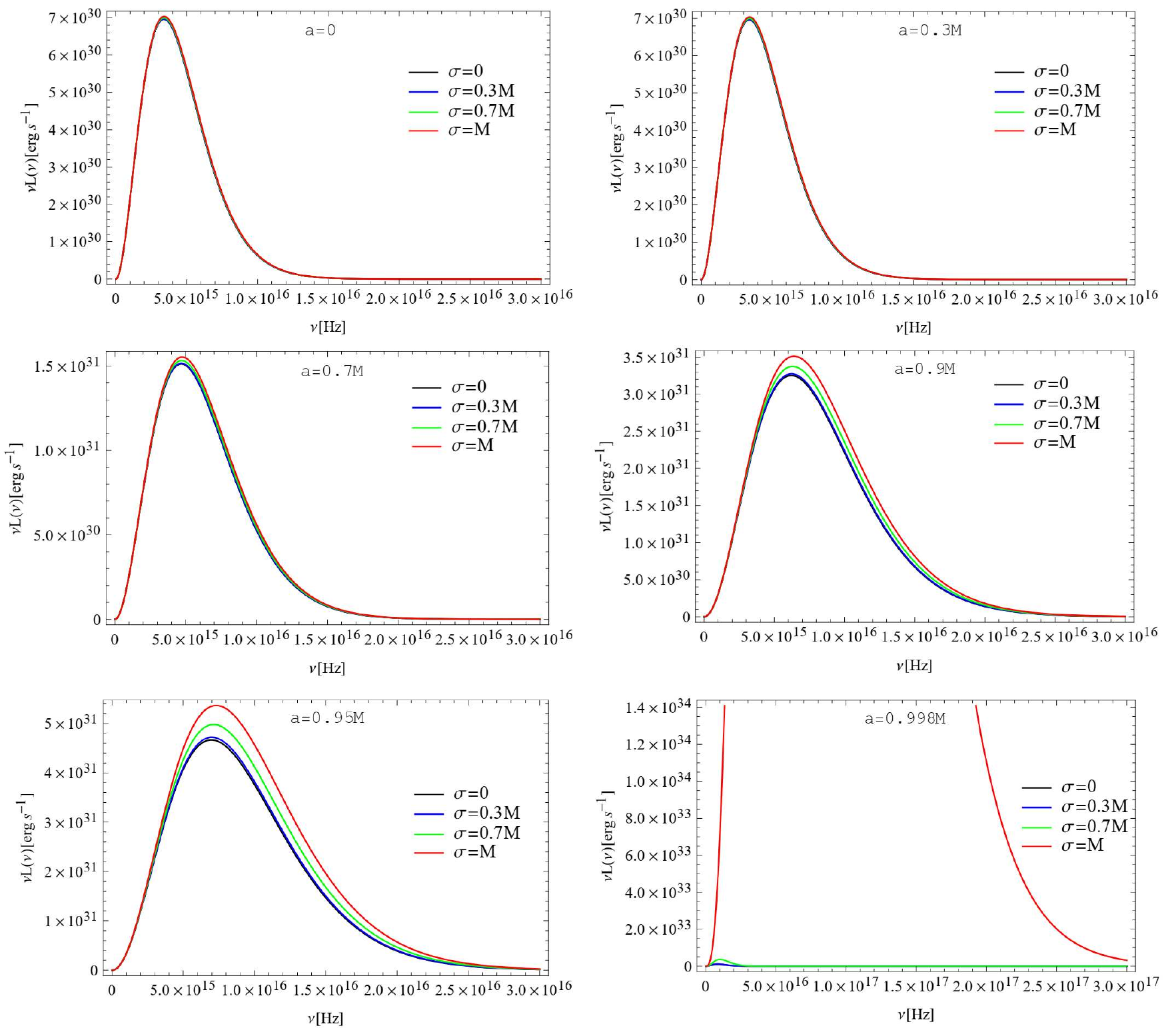}}
  \caption{Variation of luminosity $\nu L(\nu)$ in [erg.s$^{-1}$] with frequency in $\nu$[Hz].}
  \label{Lum}
\end{figure*}

Note that at the extreme spin value $a/M=1$, the efficiency $\epsilon$ becomes $42\%$. But this extreme spin value is unreachable by a realistic Kerr black hole, so we stop at the Thorne limiting value $a/M=0.998$. In Table 2, we show the maximum values of the radiation profile characteristics.

It is evident from the table that, with increase in spin $a$, the radii $r_{\scriptsize{\textmd{ms}}}$ decrease but always remain greater than the singular radius $r_{\scriptsize{\textmd{S}}}$, coming closest to $r_{\scriptsize{\textmd{S}}}$ as the spin value approaches the Thorne limit, $a/M=0.998$. So near to this limit, and with increase in scalar charge $\sigma$, we expect higher emissivity profiles in the vicinity of the singularity than those at lower values of $a/M$. Indeed, we shall see that this is precisely the case despite some metric functions and the efficiency $\epsilon$ being the same between the Bogush-Galt'sov singularity and Kerr black hole.

The most important cause responsible for heightening the maxima over the Kerr value is that the factor $\sqrt{-g}$, that appears in the denominator of flux $F(r)$ in Eq.(13), gets shrunk. The equatorial value of $\sqrt{-g}$ for the Bogush-Galt'sov singularity is \textit{not} the same as that of Kerr. For Kerr black hole, $\left. \sqrt{-g}\right\vert _{\theta = \pi /2}^{\scriptsize{\textmd{Kerr}}} = r^{2}$, while for the Bogush-Galt'sov singularity, with $b^{2} = M^{2}-a^{2}>0$, $\Delta = r^{2}-2Mr+a^{2}$, it is given by

\begin{eqnarray}
\left. \sqrt{-g}\right\vert _{\theta =\pi /2}^{\scriptsize{\textmd{BG}}} &=&r^{2}\left[1+\frac{b^{2}}{\Delta }\right] ^{-\sigma ^{2}/b^{2}}  \nonumber \\
&=&r^{2}\left[ \frac{\Delta }{\Delta +b^{2}}\right] ^{\sigma ^{2}/b^{2}}.
\end{eqnarray}%
This is much smaller than $\left. \sqrt{-g}\right\vert _{\theta =\pi /2}^{\scriptsize{\textmd{Kerr}}}$ at the ISCO radius $r=r_{\scriptsize{\textmd{ms}}}$ To get a numerical estimate of how smaller, consider the values $a=0.998M$, $r=r_{\scriptsize{\textmd{ms}}}=1.23M$ and scalar charge $\sigma =M$ (say), then
\begin{equation}
\left. \sqrt{-g}\right\vert _{\theta =\pi /2}^{\scriptsize{\textmd{BG}}}=2.90\times
10^{-9}\times \left. \sqrt{-g}\right\vert _{\theta =\pi /2}^{\scriptsize{\textmd{Kerr}}}.
\end{equation}%
Therefore, the BG profiles peak near the inner radius much higher than those of the Kerr black hole mainly because of the shrinking of $\sqrt{-g}$ as shown above. Keeping this factor in mind, we now turn to analyzing radiation profiles pointing out how other factors like $a/M$ and $\sigma$ influence the profiles.

\section{Analyses of profiles}

\subsection{Radiation flux}

Figs.1 show the variation of time averaged flux of radiation $F(r)$ with distance ($r/r_{\scriptsize{\textmd{ms}}}$) from the inner boundary of the disk $r = r_{\scriptsize{\textmd{ms}}}$ outwards with increasing order of both scalar charge $\sigma/M\leq 1$ and spin $a/M\leq 0.998$. We shall mainly focus on the profiles at $\sigma/M\sim 1$. A generic qualitative feature of the plots with increasing values of $a$ and $\sigma$ is that the profiles for BG naked singularity are always higher than those of a Kerr black hole with the BG maxima always approaching the inner radius with the increase of spin and charge. All the plots merge in the limit $r \rightarrow \infty$ since both the Kerr and BG spacetimes are asymptotically flat with the BG scalar field $\phi$ becoming trivial there.

Fig.1a (\textit{top left hand corner}) compares the flux profile for the Schwarzschild black hole ($a=0,\sigma =0$) and static naked singularities ($a=0,\sigma \neq 0$), the latter representing classes of Fischer-Janis-Newman-Winnicour (FJNW) solutions \cite{Janis:1968,Fisher:1948} (see also, \cite{Bhadra:2001b}). Both profiles peak near the radius $r\sim 1.5r_{\scriptsize{\textmd{ms}}}$, but they are almost indistinguishable. Fig.1b (\textit{top left hand corner}) compares the flux profiles of Kerr black hole ($a/M=0.3,\sigma =0$) with the spinning BG naked singularity ($a/M=0.3,\sigma /M\leq 1$). As we see, for low spin such as $a/M=0.3$, the profiles peak around $r\sim 1.5r_{\scriptsize{\textmd{ms}}}$ but again do not show any appreciable difference in $F_{\scriptsize{\textmd{max}}}(r)$.

From Table 2, we can see the exact ratios of the flux emissivity maxima $F_{\scriptsize{\textmd{max}}}(r)$ as both spin and charge increase. Fig.1c (\textit{middle left}) shows that the spin effect ($a/M=0.7$) is barely appreciable, the ratios of fluxes become $F_{\scriptsize{\textmd{max}}}^{\scriptsize{\textmd{BG}}}(r)/F_{\scriptsize{\textmd{max}}}^{\scriptsize{\textmd{Kerr}}}(r)=1.07$, which shows only a slight increase from the Kerr values. Fig.1d (\textit{middle right}) shows that the higher spin effect ($a/M=0.9$) beginning to become noticeable - the flux ratios become $F_{\scriptsize{\textmd{max}}}^{\scriptsize{\textmd{BG}}}(r)/ F_{\scriptsize{\textmd{max}}}^{\scriptsize{\textmd{ Kerr}}}(r)=1.25$. Similarly, Fig.1e (\textit{bottom left}) shows the spin effect ($a/M=0.95$) causes the flux ratio to be $F_{\scriptsize{\textmd{max}}}^{\scriptsize{\textmd{BG}}}(r)/F_{\scriptsize{\textmd{max}}}^{\scriptsize{\textmd{Kerr}}}(r)=1.56$. Profile patterns in Fig.1f (\textit{bottom right}) for ($a/M=0.998$) and ($\sigma /M=1$) show that the radiation flux from the BG singularity attains a huge value, that is, $F_{\scriptsize{\textmd{max}}}^{\scriptsize{\textmd{BG}}}(r)/F_{\scriptsize{\textmd{max}}}^{\scriptsize{\textmd{Kerr}}}(r)=4.48\times 10^{5}$ (as can be read off from Table 2), with $F_{\scriptsize{\textmd{max}}}^{\scriptsize{\textmd{BG}}}(r)$ occurring at a radius close to $r/r_{\scriptsize{\textmd{ms}}}\sim 1$, meaning that much of the radiation is emerging from the vicinity of the inner radius $r_{\scriptsize{\textmd{ms}}}$.

Thus, it turns out that the flux behavior of naked singularity becomes extremely sensitive at the Thorne spin limit $a/M$ as well as to the high values of the scalar charge $\sigma$. The exact quantitative ratios show how the radiation flux behavior of BG naked singularity could be drastically higher than those of a Kerr black hole. This feature is consistently shared by other profiles too as we shall see.

\subsection{Temperature profiles}

The qualitative features of the temperature profiles are determined by radiation flux $F(r)$ because of the Stefan-Bolzmann law, Eq.(14). Figs.2
show the variation of temperature $T(r)$ in Kelvin from the inner boundary of the disk $r=r_{\scriptsize{\textmd{ms}}}$ outwards with increasing orders of scalar charge $\sigma /M\leq 1$ and spin $a/M\leq 0.998$. As revealed from Figs.1-f, the BG profile of temperature is always higher than those of Kerr black hole, with the maxima of temperature $T_{\scriptsize{\textmd{max}}}^{\scriptsize{\textmd{BG}}}(r)$ occurring at a radius shifting toward the inner boundary of the disk, $r/r_{\scriptsize{\textmd{ms}}}=1$, with the increase of spin and scalar charge. However, as revealed from Table 2, the order of magnitude of the ratio $T_{\scriptsize{\textmd{max}}}^{\scriptsize{\textmd{BG}}}(r)/T_{\scriptsize{\textmd{max}}}^{\scriptsize{\textmd{Kerr}}}(r)\sim O(1)$ until at ($a/M=0.998$) and ($\sigma /M=1$), when $T_{\scriptsize{\textmd{max}}}^{\scriptsize{\textmd{BG}}}(r)/T_{\scriptsize{\textmd{max}}}^{\scriptsize{\textmd{ Kerr}}}(r)\sim 2.58\times 10^{2}$, which shows considerably higher temperature of the BG singularity. Thus the singularity will be much hotter than the Kerr black hole near the inner boundary of the disk consistent with the behavior of flux profiles.

\subsection{Luminosity profiles}

Figs.3 show disk luminosity spectra $\nu L(\nu )$ plotted against the observed frequency $\nu$ [Hz]. Plots show a steep rise in $\nu L(\nu )$ to $\sim 10^{30}$ [erg.s$^{-1}]$ occurring around $\nu \sim 10^{15}$ [Hz]. As the frequency $\nu$ increases, there is a steep fall to near zero luminosity. The concentration of luminosity in the narrow band of frequency width suggests that the disk appears brightest when observed within that bandwidth. The maximum luminosity ratio $\frac{\nu L(\nu )^{\scriptsize{\textmd{BG}}}}{\nu L(\nu )^{\scriptsize{\textmd{Kerr}}}}=$ $1.65\times 10^{3}$, for $a/M=0.998$ and $\sigma/M = 1$, shows that the BG singularity brightness is over $1000$ times more than that of the Kerr black hole, which is a definite signature of naked singularity.

\subsection{Differential luminosity}

Joshi et al \cite{Joshi:2014} defined a differential luminosity reaching an observer at infinity as
\begin{equation}
\frac{d\mathcal{L}_{\infty }}{d\ln {r}}=4\pi r\sqrt{-g}\widetilde{E}F,
\end{equation}%
where $\widetilde{E}$ is the specific energy and $F$ is the radiation flux emitted by the innermost radius of the disk. Note that $\sqrt{-g}$ cancels with the $1/\sqrt{-g}$ in the definition of $F$ leaving only%
\begin{equation}
\frac{d\mathcal{L}_{\infty }}{d\ln {r}}=r\widetilde{E}_{\scriptsize{\textmd{ms}}}\times
\frac{-\dot{M}_{0}\Omega _{,r}}{\left( \widetilde{E}-\Omega \widetilde{L}%
\right) ^{2}}\int_{r_{\scriptsize{\textmd{ms}}}}^{r}\left( \widetilde{E}-\Omega \widetilde{L%
}\right) \widetilde{L}_{,r}dr
\end{equation}

As noted before, $\widetilde{E}(r_{\scriptsize{\textmd{ms}}})$ for the BG is the same as that of Kerr black hole having a maximum value $0.679$ and the multiplying factor being finite for $a\leq 0.998M$, the differential luminosity reaching an observer at the asymptotically flat region increases without limit.

\section{Conclusions}
\label{sec:4}

We presented a comparative study of accretion profiles between Bogush-Gal'tsov spinning naked singularity of general relativity (GR) and Kerr black hole. Our main conclusion is that the emissivity profiles of naked singularity peak much higher near the ISCO boundary than the ones belonging to Kerr black hole.

To reach our conclusion, we had worked within the framework of Novikov-Thorne model of thin accretion disk around a stellar sized spinning object. This model is extremely useful and widely employed to study accretion phenomena. The motivation of our work was to compare, both qualitatively and quantitatively, the discriminating emissivity features between a GR spinning naked singularity and Kerr black hole. Our study contrasts with a previous work in \cite{Shahidi:2020} that compared Novikov-Thorne emissivity properties of Brans-Dicke (BD) spinning naked singularity and Kerr black hole.

Within the realm of BD theory, an early spinning solution was generated by Kim \cite{Kim:1999}, where he found that nontrivial Kerr-Newman-type black hole solutions (hairy spinning black holes) \textit{different} from general relativistic solutions could occur for the Brans-Dicke parameter values $-5/2\leq \omega _{\scriptsize{\textmd{BD}}}<-3/2$. Its implications for the Penrose conjecture was also discussed in \cite{Kim:1999}. The non-spinning uncharged corollary yields static but non-spherically symmetric hairy black hole and not the usual Schwarzschild black hole of general relativity. Defining the solution parameter as $\left\vert 1-\gamma \right\vert =\frac{4}{2\omega _{\scriptsize{\textmd{BD}}}+3}$, Shahidi, Harko and Kov\'{a}cs (SHK) \cite{Shahidi:2020} classified the Kim solution into three categories: Brans-Dicke-Kerr model of naked singularity ($-\infty <\gamma \leq 0$ and $2<\gamma <\infty $), non-general-relativistic hairy spinning black hole ($0<\gamma <1$ and $1<\gamma <2$) and the usual Kerr black hole ($\gamma =1$).

On the other hand, within the realm of GR theory, the latest spinning naked singularity obtained by Bogush and Galt'sov (BG) \cite{Bogush:2020} in the Einstein-scalar system directly reduces to Kerr black hole when the scalar field is switched off. Thus, spinning naked singularity in GR is different from the same in BD theory treated in \cite{Shahidi:2020} in the sense that the former does not yield hairy spinning black holes like the ones in \cite{Kim:1999} but directly yields only the hairless Kerr black hole as a special case ($\sigma =0$).

It is interesting to note that the frame dragging frequency $\omega_{\scriptsize{\textmd{fd}}} = -\frac{g_{t\varphi}}{g_{\varphi\varphi}}$ and the conversion efficiency $\epsilon$ in the two models of spinning naked singularities, viz., in the BD and the GR models, are exactly the same as those of the Kerr black hole. There is no \textit{a priori} reason that the metric components $g_{t\varphi}$ and $g_{\varphi\varphi}$ in two distinct gravity theories be the same, while other metric components differ. The conversion efficiency $\epsilon$ of the BG naked singularity is \textit{independent} of $\sigma$ and hence at the Thorne limit, it is the same as the Kerr value $32.1\%$ (Table 1). It has also been found that, on the equatorial plane ($\theta = \pi/2$), kinematic properties of BG naked singularity are exactly the same as those of Kerr black hole. Therefore, none of the above properties (frame dragging, efficiency, kinematic properties) can distinguish BD and GR models of naked singularities from their limiting Kerr black hole counterpart. Thus, looking for distinctive features, we studied the Novikov-Thorne emissivity profiles such as the flux, temperature and luminosity, assuming a toy model of stellar sized BG spinning naked singularity. We assumed, for a meaningful comparison, that the BG naked singularity respects the Thorne limit on the black hole spin, $a/M\leq 0.998$.

The most illuminating results from the present work are the \textit{quantitative} predictions of the maxima of the emissivity profiles displayed in Table 2. As shown there, the accretion disk emission profiles of BG spinning singularity could be much higher than those of a Kerr black hole - naked singularities are hotter and brighter as the spin $a$ and scalar charge $\sigma$ increase. In particular, the radiative flux could be considerably higher near the inner disk boundary, e.g., $10^{5}$ times higherr than that of a Kerr black hole under the same spin values but non-trivial scalar charges. This happens largely due to the fact that near the inner boundary $\sqrt{-g}$ considerably shrinks compared to the Kerr value as shown in Eq.(30). We speculate that such drastically higher profiles of naked singularity could actually be its generic feature as similar features are exhibited also by other entirely different models of naked singularity such as the Brans-Dicke-Kerr model \cite{Kim:1999,Shahidi:2020} and the JMN perfect fluid model \cite{Joshi:2014}.

We point out a recent work dealing with experimental relativity with accretion disk observations \cite{Cardenas:2019}. The authors, using the Rezzolla-Zhidenko \cite{Rezzolla:2014} bumpy black hole, combined ray-tracing and MCMC sampling techniques to constrain and detect deviations from general relativity using the accretion disk spectrum of stellar-mass black holes in binary systems. They concluded that, "even when a very simple astrophysical model for the accretion disk is assumed a priori, the uncertainties and covariances between the parameters of the model and the parameters that control the deformation from general relativity make any test of general relativity very challenging with accretion disk spectrum observations." As to our metric, however, there is no deformation from general relativity but there is a scalar hair $\sigma$ causing naked singularity and statistical techniques using observed accretion data can, in principle, constrain $\sigma$, which would have implications especially for the Ruffini-Wheeler conjecture on hairless black holes. This topic however would be beyond the motivation of the present work but certainly worth pursuing in the future.

\appendix

\section{Axially symmetric spacetime}

The BG spacetime is part of one of the axially symmetric spacetime parametrisations suggested in \cite{Konoplya:2016}:
\begin{eqnarray}
d\tau^{2} &=& - \frac{N^{2}(r, \theta) - W^{2}(r, \theta)}{K^{2}(r, \theta)}dt^{2} - 2W(r, \theta)r\sin^{2}{\theta}dtd\phi \nonumber \\
&& + K^{2}(r, \theta)r^{2}\sin^{2}{\theta}d\varphi^{2}  \nonumber \\
&& + \Lambda(r, \theta)\left[\frac{B^{2}(r, \theta)}{N^{2}(r, \theta)}dr^{2} + r^{2}d\theta^{2}\right].
\end{eqnarray}%
The\ BG metric part can be rewritten as%
\begin{eqnarray}
d\tau_{\scriptsize{\textmd{BG}}}^{2} &=& -\left(1 - \frac{2Mr}{\Sigma}\right)dt^{2} - \frac{4aMr\sin^{2}{\theta}}{\Sigma}dtd\phi \nonumber \\
&& + \left(r^{2} + a^{2} + \frac{2a^{2}Mr\sin^{2}{\theta}}{\Sigma}\right)\sin^{2}{\theta}d\varphi^{2} \nonumber \\
&& + \left[1 + \frac{b^{2}}{\Delta}\sin^{2}\theta\right]^{-\sigma^{2}/b^{2}}\Sigma\left[\frac{dr^{2}}{\Delta} + d\theta^{2}\right].
\end{eqnarray}%
The functions in (A.1) can be identified from the BG solution as follows
\begin{eqnarray}
N^{2}\left(r, \theta\right) &=& \frac{\left(\Sigma - 2Mr\right) \left(r^{2} + a^{2}\right) + 2a^{2}Mr\sin^{2}{\theta}}{r^{2}\Sigma}, \\
W\left(r, \theta\right) &=& \frac{2aM}{\Sigma},  \\
K^{2}\left(r, \theta\right) &=& 1 + \frac{a^{2}}{r^{2}} + \frac{2a^{2}M\sin^{2}{\theta}}{r\Sigma},  \\
\Lambda\left(r, \theta\right) &=& \frac{\Sigma}{r^{2}}\left(1 + \frac{b^{2}\sin^{2}{\theta}}{\Delta}\right)^{-\sigma^{2}/b^{2}},  \\
B^{2}\left(r, \theta\right) &=& \frac{\left(\Sigma - 2Mr\right)\left(r^{2} + a^{2}\right) + 2a^{2}Mr\sin^{2}{\theta}}{\Delta\Sigma},
\end{eqnarray}%
where
\begin{eqnarray}
\Sigma  &=& r^{2}+a^{2}\cos ^{2}{\theta },   \\
\Delta  &=& r^{2}-2Mr+a^{2},   \\
b &=& \sqrt{M^{2}-a^{2}}.
\end{eqnarray}%
Using the asymptotic behavior of the functions (A.3-A.7), one may read off the asymptotic and strong field coefficients derived in \cite{Konoplya:2016}.


\begin{thebibliography}{99}

\bibitem{Bhadra:2001a}
  A. Bhadra and K.K. Nandi, Phys. Rev. D \textbf{64}, 087501 (2001).

\bibitem{Faraoni:1999}
  V. Faraoni, Phys. Rev. D \textbf{59}, 084021 (1999).

\bibitem{Matos:2000}
  T. Matos, F.S. Guzm\'{a}n znd D. Nu\~{n}ez, Phys. Rev. D \textbf{62}, 061301 (2000).

\bibitem{Nandi:2009}
  K.K. Nandi, I. Valitov and N.G. Migranov, Phys. Rev. D \textbf{80}, 047301 (2009).

\bibitem{Tiwari:1976}
  R.N. Tiwari and B.K. Nayak, Phys. Rev. D \textbf{14}, 2502, (1976); J. Math. Phys. \textbf{18}, 289 (1977).

\bibitem{Singh:1979}
  T. Singh and L.N. Rai, Gen. Relativ. Gravit. \textbf{11}, 37 (1979).

\bibitem{Misra:1972}
  R.M. Misra and D.B. Pandey, J. Math. Phys. \textbf{13}, 1538, (1972).

\bibitem{Kim:1999}
  H. Kim, Phys. Rev. D \textbf{60}, 024001 (1999).

\bibitem{Novikov:1973}
  I.D. Novikov and K.S. Thorne, \textit{Astrophysics and black holes}, in Black Holes, edited by C. De Witt and B. De Witt (Gordon and Breach, New York, 1973).

\bibitem{Shakura:1973}
  N.I. Shakura and R.A. Sunyaev, Astron. Astrophys. \textbf{24}, 33 (1973).

\bibitem{Page:1974}
  D.N. Page and K.S. Thorne, Astrophys. J. \textbf{191}, 499 (1974).

\bibitem{Thorne:1974}
  K.S. Thorne, Astrophys. J \textbf{191}, 507 (1974).

\bibitem{Shahidi:2020}
  S. Shahidi, T. Harko and Z. Kov\'{a}cs, Eur. Phys. J. C \textbf{80}, 162 (2020).

\bibitem{Bogush:2020}
  I. Bogush and D. Galt'sov, Phys. Rev. D \textbf{102}, 124006 (2020).

\bibitem{Penrose:1969}
  R. Penrose, Riv. Nuovo Cim. \textbf{1}, 252 (1969).

\bibitem{Ruffini:1971}
  R. Ruffini and J.A. Wheeler, Phys. Today \textbf{24}, 30 (1971).

\bibitem{Joshi:2007}
  P.S. Joshi, \textit{Gravitational Collapse and Spacetime Singularities} (Cambridge University Press, Cambridge, 2007).

\bibitem{Joshi:2014}
  P.S. Joshi, D. Malafarina and R. Narayan, Class. Quantum Grav. \textbf{31}, 015002 (2014).

\bibitem{Janis:1968}
  A.I. Janis, E.T. Newman and J. Winnicour, Phys. Rev. Lett. \textbf{20}, 878 (1968).

\bibitem{Bhadra:2001b}
  A. Bhadra and K.K. Nandi, Int. J. Mod. Phys. A \textbf{28}, 4543 (2001).

\bibitem{Agnese:1995}
  A.G. Agnese and M. La Camera, Phys. Rev. D \textbf{51}, 2011 (1995).

\bibitem{Nandi:1997}
  K.K. Nandi, A. Islam and J. Evans, Phys. Rev. D \textbf{55}, 2497 (1997).

\bibitem{Anchordoqui:1997}
  L.A. Anchordoqui, S.P. Bergliaffa and D.F. Torres, Phys. Rev. D \textbf{55}, 526 (1997).

\bibitem{Nandi:1998}
  K.K. Nandi, B. Bhattacharjee, S.M.K. Alam and J. Evans, Phys. Rev. D \textbf{57}, 823 (1998).

\bibitem{Nandi:2004}
  K.K. Nandi and Y.-Z. Zhang, Phys. Rev. D\textbf{\ 70}, 044040 (2004).

\bibitem{Bhattacharya:2009}
  A. Bhattacharya, I. Nigmatzyanov, R. Izmailov and K.K. Nandi, Class. Quantum Grav. \textbf{26}, 235017 (2009).

\bibitem{Cardoso:2016}
  V. Cardoso, E. Franzin and P. Pani, Phys. Rev. Lett. \textbf{116}, 171101 (2016).

\bibitem{Nandi:2006}
  K.K. Nandi, Y.-Z. Zhang and A.V. Zakharov, Phys. Rev. D \textbf{74}, 024020 (2006).

\bibitem{Nandi:2017}
  K.K. Nandi, R.N. Izmailov, A.A. Yanbekov and A.A. Shayakhmetov, Phys. Rev. D \textbf{95}, 104011 (2017).

\bibitem{Lukmanova:2018}
  R.F. Lukmanova, G.Y. Tuleganova, R.N. Izmailov and K.K. Nandi, Phys. Rev. D \textbf{97}, 124027 (2018).

\bibitem{Virbhadra:2002}
  K.S. Virbhadra and G.F.R. Ellis, Phys. Rev. D \textbf{65}, 103004 (2002).

\bibitem{Izmailov:2019}
  R.N. Izmailov, R.Kh. Karimov, E.R. Zhdanov and K.K. Nandi, Mon. Not. R. Astron. Soc. \textbf{483}, 3754 (2019).

\bibitem{Tsukamoto:2017}
  N. Tsukamoto and Y. Gong, Phys . Rev . D \textbf{95} , 064034 (2017).

\bibitem{Shaikh:2017}
  R. Shaikh and S. Kar, Phys. Rev. D \textbf{96}, 044037 (2017).

\bibitem{Nakajima:2012}
  K. Nakajima and H. Asada, Phys. Rev. D \textbf{85}, 107501 (2012).

\bibitem{Shaikh:2019}
  R. Shaikh, P. Kocherlakota, R. Narayan and P.S. Joshi, Mon. Not. R. Astron. Soc. \textbf{482}, 52 (2019).

\bibitem{Shaikh:2018}
  R. Shaikh, Phys. Rev. D \textbf{98}, 024044 (2018).

\bibitem{Longair:1994}
  M.S. Longair, \textit{High Energy Astrophysics, Vol. II} (Cambridge University Press, Cambridge, 1994).

\bibitem{Torres:2002}
  D. Torres, Nucl. Phys. B \textbf{626}, 377 (2002).

\bibitem{Bambi:2013}
  C. Bambi and D. Malafarina, Phys. Rev. D \textbf{88}, 064022 (2013).

\bibitem{Karimov:2019}
  R.Kh. Karimov, R.N. Izmailov and K.K. Nandi, Eur. Phys. J. C \textbf{79}, 952 (2019).

\bibitem{Harko:2009a}
  T. Harko, Z. Kov{\'{a}}cs and F.S.N. Lobo, Class. Quantum Grav. \textbf{26}, 215006 (2009).

\bibitem{Harko:2009b}
  T. Harko, Z. Kov{\'{a}}cs and F.S.N. Lobo, Phys. Rev. D \textbf{79}, 064001 (2009).

\bibitem{Karimov:2020}
  R.Kh. Karimov, R.N. Izmailov, A.A. Potapov and K.K. Nandi, Eur. Phys. J. C \textbf{80}, 1138 (2020).

\bibitem{Yusupova:2021}
  R.M. Yusupova, R. Kh. Karimov, R.N. Izmailov and K.K. Nandi, Universe \textbf{7}, 177 (2021).

\bibitem{Heydari-Fard:2021}
  M. Heydari-Fard, M. Heydari-Fard and H.R. Sepangi, Eur. Phys. J. C  \textbf{81}, 473 (2021).

\bibitem{Thorne:1972}
  K.S. Thorne, in J. Klauder (Ed.) \textit{Magic Without Magic: John Archibald Wheeler} (Freeman, San Francisco, 1972).

\bibitem{Hod:2018}
  S. Hod, Eur. Phys. J. C \textbf{78}, 1013 (2018).

\bibitem{Hod:2020a}
  S. Hod, Eur. Phys. J. C \textbf{80}, 162 (2020).

\bibitem{Hod:2020b}
  S. Hod, Eur. Phys. J. C \textbf{80}, 1148 (2020).

\bibitem{Nandi:2020}
  K.K. Nandi, R.N.Izmailov, G.M.Garipova, R.R. Volotskova and A.A. Potapov, Phys. Lett. B \textbf{809}, 135734 (2020).

\bibitem{Nandi:2021}
  K.K. Nandi, R.N.Izmailov, A.A. Potapov, K.K. Nandi and N.G. Migranov, Eur. Phys. J. C \textbf{81}, 997 (2021).

\bibitem{Bronnikov:2011}
  K.A. Bronnikov, J.C. Fabris and A. Zhidenko, Eur. Phys. J. C \textbf{71}, 1791 (2011).

\bibitem{Bardeen:1972}
  J.H. Bardeen, W.H. Press and S.A. Teukolsky, Astrophys. J. \textbf{178}, 347 (1972).

\bibitem{Chavineau:2019}
  B. Chavineau, Phys. Rev. D \textbf{100}, 024051 (2019).

\bibitem{Fisher:1948}
  I.Z. Fisher, Zh. Eksp. Teor. Fiz. \textbf{8}, 636 (1948).

\bibitem{Cardenas:2019}
  A. C\'{a}rdenas-Avenda\~{n}o et al, Phys. Rev. D \textbf{100}, 024039 (2019).

\bibitem{Rezzolla:2014}
  L. Rezzolla and A. Zhidenko, Phys. Rev. D \textbf{90}, 084009 (2014).

\bibitem{Konoplya:2016}
  R. Konoplya, L. Rezzolla and A. Zhidenko, Phys. Rev. D \textbf{93}, 064015 (2016).
\end{thebibliography}
\end{document}